\documentclass[nature,superscriptaddress]{revtex4}
\usepackage{comment}
\usepackage{amsmath}
\usepackage{amsfonts}
\usepackage{amssymb}
\usepackage{mathrsfs}
\usepackage{graphics}
\usepackage[dvips]{graphicx}
\newcommand{\ket}[1]{\left | #1 \right\rangle}

\renewcommand{\epsilon}{\varepsilon}

\begin{document}

\bibliographystyle{naturemag}

\title{Entanglement between superconducting qubits and a tardigrade}

\author{K. S. Lee}
\affiliation{School of Physical and Mathematical Sciences, Nanyang Technological University, Singapore}

\author{Y. P. Tan}
\affiliation{School of Physical and Mathematical Sciences, Nanyang Technological University, Singapore}

\author{L. H. Nguyen}
\affiliation{School of Physical and Mathematical Sciences, Nanyang Technological University, Singapore}

\author{R. P. Budoyo}
\affiliation{Centre for Quantum Technologies, National University of Singapore, Singapore}

\author{K. H. Park}
\affiliation{Centre for Quantum Technologies, National University of Singapore, Singapore}

\author{C. Hufnagel}
\affiliation{Centre for Quantum Technologies, National University of Singapore, Singapore}

\author{Y.~S.~Yap}
\affiliation{Faculty of Science and Centre for Sustainable Nanomaterials (CSNano), Universiti Teknologi Malaysia, 81310 UTM Johor Bahru, Johor, Malaysia}
\affiliation{Centre for Quantum Technologies, National University of Singapore, Singapore}

\author{N. M{\o}bjerg}
\affiliation{Department of Biology, University of Copenhagen, Denmark}

\author{V. Vedral}
\affiliation{Department of Physics, University of Oxford, United Kingdom}
\affiliation{Centre for Quantum Technologies, National University of Singapore, Singapore}
\affiliation{Department of Physics, National University of Singapore, Singapore}

\author{T. Paterek}
\affiliation{Institute of Theoretical Physics and Astrophysics, University of Gda{\'n}sk, Poland}

\author{R. Dumke}
\affiliation{School of Physical and Mathematical Sciences, Nanyang Technological University, Singapore}
\affiliation{Centre for Quantum Technologies, National University of Singapore, Singapore}

\begin{abstract}
Quantum and biological systems are seldom discussed together as they seemingly demand opposing conditions.
Life is complex, ``hot and wet'' whereas quantum objects are small, cold and well controlled.
Here, we overcome this barrier with a tardigrade --- a microscopic multicellular organism known to tolerate extreme physiochemical conditions via a latent state of life known as cryptobiosis. We observe coupling between the animal in cryptobiosis and a superconducting quantum bit and prepare a highly entangled state between this combined system and another qubit.
The tardigrade itself is shown to be entangled with the remaining subsystems.
The animal is then observed to return to its active form after 420 hours at sub 10 mK temperatures and pressure of $6\times 10^{-6}$ mbar, setting a new record for the conditions that a complex form of life can survive.
\end{abstract}

\maketitle

In his highly influential 1933 lectures titled “Light and Life”~\cite{BOHR1933} Niels Bohr espoused an analogy between physicists’ attempt to characterize the atom and biologists’ attempt to characterize the cell. Living cells are made of ordinary matter, and therefore fully subject to chemical analysis, but the matter is organized in a complex and intricate way. Bohr proposed that in order to study the chemistry of the cell, the organization had to be destroyed. On the other hand, to study the organization, one has to operate at a level at which the chemistry is invisible. Bohr therefore proposed that the chemical basis of an organism and its organizational hierarchy are complementary properties, and could never be studied simultaneously.

Here we not only go beyond Bohr’s complementarity by entangling a living system to two superconducting quantum bits, but we also expose an additional twist in areas where quantum physics, chemistry and biology all overlap.  Namely, the living system we study is a tardigrade which is capable of suspending its living functions for an indefinite amount of time if the external conditions are deemed to be adverse enough for survival~\cite{moberg2021}. In that sense, and this is a point that Bohr certainly missed, it is possible to do a quantum and hence a chemical study of a system, without destroying its ability to function biologically. 

Our experiments could be classified within the field of quantum biology~\cite{Lambert2012} which studies the possibility of quantum effects in biologically-relevant processes such as photosynthesis~\cite{Cao2020}, olfaction~\cite{Brookes2017} or animal navigation~\cite{Wiltschko2019}, but is also concerned with quantum features of entire organisms. Along this direction, a systematic body of work with matter-wave interferometers revealed wave properties of molecules with growing complexity and holds promise for similar outcomes measured with viruses~\cite{Arndt1999,Hackermller2003,Gerlich2011}.
Living photosynthetic bacteria have been observed to strongly couple to light~\cite{Coles2017} and under certain assumptions the results obtained could be interpreted in terms of entanglement between the quantised light and light-sensitive parts of bacteria~\cite{Marletto2018}.
Our experiments involve a multi-cellular eukaryote that is a much more complex unit of life interacting electrically with quantum bits and for which conclusions about quantum features are unavoidable.

In our experiments, we use specimens of a Danish population of \textit{Ramazzottius varieornatus} Bertolani and Kinchin, 1993 (Eutardigrada, Ramazzottiidae). The species belongs to phylum Tardigrada comprising of microscopic invertebrate animals with an adult length of 50-1200 $\mu$m~\cite{moberg2018_2}. Importantly, many tardigrades show extraordinary survival capabilities~\cite{Mobjerg11} and selected species have previously been exposed to extremely low temperatures of 50 mK \cite{Paul1950} and low Earth orbit pressures of $10^{-19}$ mbar~\cite{Jnsson2008}.
Their survival in these extreme conditions is possible thanks to a latent state of life known as cryptobiosis \cite{moberg2021,Mobjerg11}. Cryptobiosis can be induced by various extreme physicochemical conditions, including freezing and desiccation. Specifically, during desiccation, tardigrades reduce volume and contract into an ametabolic state, known as a ``tun". Revival is achieved by reintroducing the tardigrade into liquid water at atmospheric pressure. In the current experiments, we used dessicated \textit{R. varieornatus} tuns with a length of 100-150 $\mu$m. Active adult specimens have a length of 200-450 $\mu$m. The revival process typically takes several minutes.

We place a tardigrade tun on a superconducting transmon qubit 
and observe coupling between the qubit and the tardigrade tun via a shift in the resonance frequency of the new qubit-tardigrade system.
This joint qubit-tardigrade system is then entangled with a second superconducting qubit. We reconstruct the density matrix of this coupled system experimentally via quantum state tomography.
Finally, the tardigrade is removed from the superconducting qubit and reintroduced to atmospheric pressure and room temperature. We observe the resumption of its active metabolic state in water. Notably, the tardigrade tun remained at a base temperature below 10 mK and pressures around $6\times10^{-6}$ mbar for 420 hours (see Supplementary Information (SI) for more details). This is to-date the most extreme exposure to low temperatures and pressures that a tardigrade has been recorded to survive, clearly demonstrating that the state of cryptobiosis ultimately involves a suspension of all metabolic processes~\cite{Keilin,clegg2001} given that all chemical reactions would be prohibited with all its constituent molecules cooled to their ground states.

\begin{figure*}[tpb]
\includegraphics[width=\textwidth]{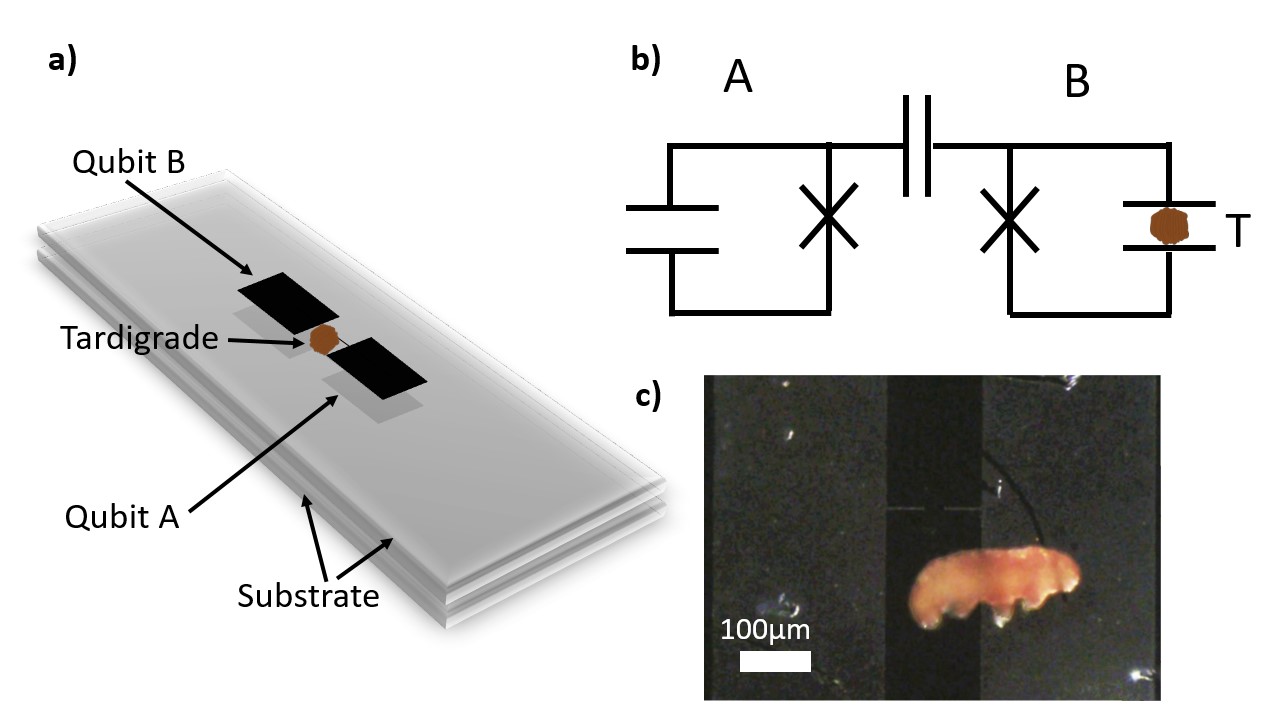}
\caption{
\label{FIG_SCHEME} Sketch of the experiment. a)
A tardigrade in the tun state is positioned between the shunt capacitior plates, slightly offset from the Josephson junction, of the transmon qubit B. Qubit A is on the underside and is capacitively coupled to qubit B. The full chip is placed within a 3D copper cavity that is mounted inside a dilution refrigerator and connected to standard microwave electronics for probing. b) Circuit diagram of the two qubits and the tardigrade tun. c) Magnification of the revived tardigrade on transmon qubit. The tardigrade in the tun state is placed in the same position  while still attached to a small piece of filter paper during the experiment.
}
\end{figure*}

Figure~\ref{FIG_SCHEME} shows an illustration of the experiment.
Two transmon qubits (`qubit A' and `qubit B') are mounted back to back inside an oxygen-free high thermal-conductivity copper 3D microwave cavity with a TE$_{101}$ frequency of 4.521~GHz.
The two transmons are positioned about 1~mm apart (twice the thickness of the silicon chips). 
The two qubits are directly coupled via the shunt capacitor plates on each qubit.
We characterize the system using control and readout schemes typically used in superconducting qubits~\cite{Lin2021}.
The frequency of each qubit is tunable, but is always biased close to the maximum in our experiments.
These correspond to $f_A=3.048$~GHz for qubit A and $f_B=3.271$~GHz for qubit B without tardigrade tun.

\begin{figure*}[tpb]
\includegraphics[width=\textwidth]{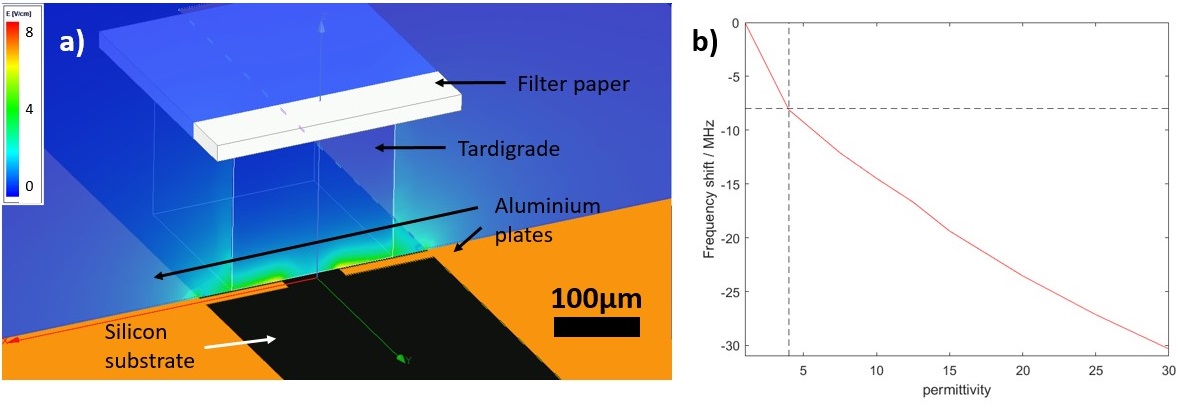}
\caption{
\label{FIG_RESULTS} a) Numerical simulations of the electric field along the surface of the tardigrade for an exemplary potential difference of 20 mV across the shunt capacitor plates (shown in orange). The tardigrade in the tun state was simulated as a cube of length 100~$\mu$m, with dielectric constant of $4$, attached to a 150~$\mu$m square filter paper with dielectric constant of 4.5. As seen, the filter paper is sufficiently displaced from the transmon plane and does not interact significantly with the electric fields. b) Frequency shift expected from tardigrade, simulated with typical permittivity values for proteins (in the range 4 to 30).  The observed frequency shift of -8 MHz corresponds to a dielectric constant of about 4 for the tardigrade tun.}
\end{figure*}

For the subsequent cooldown, a single cryptobiotic tardigrade on a 150~$\mu$m square piece of cellulose filter paper is placed on the substrate of qubit B, in between the shunt capacitor plates of the transmon qubit, with the tardigrade side closer to the substrate. The filter paper is sufficiently displaced from the transmon plane and interacts very weakly with the transmon electric fields, verified with numerical simulations (see Fig.~\ref{FIG_RESULTS}a).
The maximum frequency of qubit B is shifted down by 8~MHz to $f'_B=3.263$~GHz, as shown in Fig.~\ref{FIG_RESULTS}b, while the maximum frequency of qubit A did not change significantly.
The shift can be attributed to the aging of the Josephson junctions and the addition of the tardigrade itself. 
The addition of the tardigrade results in a change of the effective dielectric medium and contributes to the transmon shunt capacitance of qubit B, consequently shifting its frequency. We simulate the electric fields and capacitance shifts using ANSYS Maxwell where the tardigrade is modelled as a cube of length 100~$\mu$m positioned as in the experiment and with dielectric constant varied between 4 to 30, typical values for proteins~\cite{li2013} (the filter paper is also taken into account and verified to have small contribution). 
Fig.~\ref{FIG_RESULTS}b shows that the measured frequency shift corresponds to tardigrade dielectric constant of $\varepsilon_r\approx4$.
Note that this is in the lower range of the dielectric constants for proteins indicating that the contribution from the aging effect is small, in agreement with measurements on qubit A.

This macroscopic behaviour can be understood with a microscopic model where the charges inside the tardigrade are represented as effective harmonic oscillators that couple to the electric field of the qubit via the dipole mechanism. This is a modification of the two-level systems used to model defects in dielectric structures \cite{sarabi2016}. Due to this coupling, the combined system has different energy gap as compared to just the bare qubit.
We model the charges inside the tardigrade as a collection of $N$ harmonic oscillators (with possibly different frequencies) coupled to the qubit.
Due to the small observed energy shift it is reasonable to assume that the coupling is weak and could be treated as a perturbation.
We therefore split the Hamiltonian and write the following base term
\begin{align}
        H_0& = - \frac{1}{2} \hbar \omega_q \, \sigma_z + \hbar \sum_{j=1}^N \omega_j \, a^\dagger_j a_j,
\end{align}
where $\sigma_{z}$ is the Pauli-$z$ operator for the qubit with lower (higher) energy state denoted as $| 0 \rangle$ ($| 1 \rangle$) and $a_j$ ($a_j^\dagger$) is the lowering (raising) operator for the $j$th oscillator.
Note that vanishing energy of the qubit has been set in the middle between its energy levels and the ground state of each oscillator is assigned energy zero.
The perturbation term is given by the coupling, where again due to weak interaction we also ignore so-called counter-rotating terms capable of simultaneous excitation of the qubit and an oscillator.
Therefore,
\begin{align}
        V&=\frac{\hbar g}{2} \sum_{j=1}^N (a^\dagger_j \sigma_- + a_j \sigma_+),
\end{align}
where for simplicity the same coupling $g$ is used for each oscillator and we write $\sigma_+ = | 1 \rangle \langle 0 |$ and $\sigma_- = | 0 \rangle \langle 1 |$.
One verifies that the ground state of this system is given by $| g \rangle = | 0 \dots 0 \rangle$ for any coupling $g$, i.e. the qubit and all oscillators are in their respective ground states.
The first excited state and its energy can be obtained via the second-order perturbation theory and show the following frequency gap to the ground state:
\begin{equation}
    \Delta f = \frac{1}{2\pi}\left(\omega_q+\frac{g^2}{4}\sum_{j=1}^N \frac{1}{\omega_q-\omega_j}\right).
    \label{eq:generalshift}
\end{equation}
Accordingly, if the sum is negative, the frequency difference is lowered in the presence of interaction.
This happens when there are charges in the tardigrade that oscillate faster than the qubit frequency.
We shall also need the explicit form of the first excited state which is found to read:
\begin{equation}
    \ket{e} = \cos \frac{\theta}{2} \, \ket{1} \ket{0 \dots 0}+\sin \frac{\theta}{2} \, \ket{0}\ket{\psi_1},
    \label{EQ_E}
\end{equation}
where the first subsystem denotes the qubit and the remaining subsystems describe the oscillators.
The state $\ket{\psi_1}$ is the superposition of a single excitation in each oscillator, i.e. $\ket{\psi_1} = c_1 \ket{10\dots 0} + \dots + c_N \ket{0 \dots 01}$ with suitable coefficients,
and $\cos \frac{\theta}{2} = 1 - (g^2/8)\sum_j(1/\delta_j^2)$,
where $\delta_j = \omega_q - \omega_j$ is the detuning.
Note that this model reduces to the well-known Tavis-Cummings model in the case of identical oscillators~\cite{Tavis1968}.

\begin{figure*}[tpb]
\includegraphics[width=\linewidth]{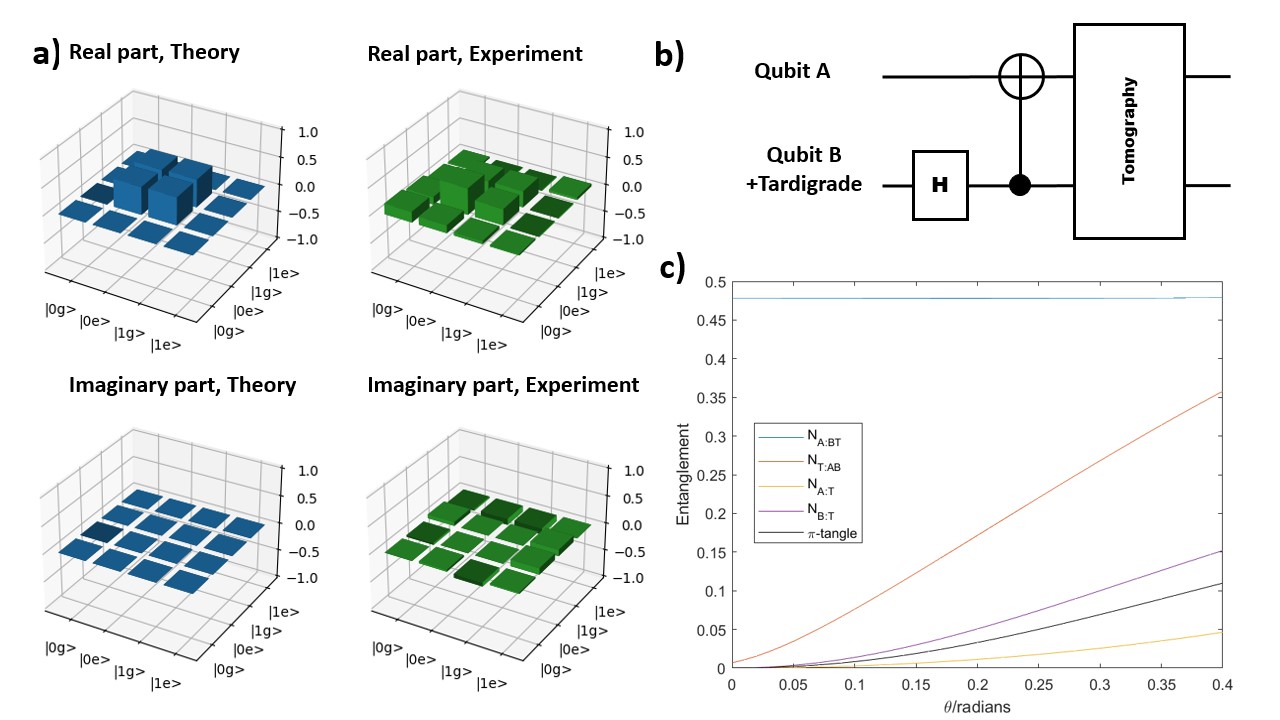}
\caption{Entanglement with the tardigrade. a) Tomography of the combined tripartite system in the four-dimensional subspace spanned by the states of qubit A and dressed states of qubit B and tardigrade tun together.
The fidelity of the density matrix reconstructed via maximum likelihood estimation (green) to the ideal state (blue) is $82 \%$. b) The Hadamard gate (H) followed by the CNOT gate was used to produce the state presented in panel a).
c) Quantum entanglement in the microscopic model as a function of $\theta$, interaction strength between the tardigrade tun and qubit B. The $\pi$-tangle quantifies genuine tripartite entanglement, i.e. a form of quantum correlations that cannot be reduced to pairwise entanglement.
See SI for formal definitions.
 }
\label{fig:tomo_entanglement}
\end{figure*}

For the second part of the experiment, we couple the qubit B-tardigrade system with qubit A using a pulse sequence shown in Fig.~\ref{fig:tomo_entanglement}. The controlled-NOT gate uses a pulse shape given by Speeding up Waveforms by Inducing Phases to Harmful Transitions (SWIPHT) protocol \cite{Economou2015, Premaratne2019}. 
This pulse sequence ideally produces the entangled state $\frac{1}{\sqrt{2}}(\ket{0_\mathrm{A}, e_{\mathrm{BT}}}+\ket{1_\mathrm{A}, g_{\mathrm{BT}}})$, where we have used subscripts to demarcate the subsystems.
To verify the entangled state, we performed quantum state tomography in the four-dimensional subspace of the whole combined tripartite system. We applied 16 different combinations of one-qubit gates on qubit A and dressed states of the joint qubit B-tardigrade system. We then jointly readout the state of both qubits using the cavity \cite{Filipp2009}.
Maximum likelihood estimation \cite{James2001} was then employed to prevent the resulting density matrix from having nonphysical properties. Comparing the experimental density matrix with the expected density matrix, we find a state fidelity of $F=82$~\% \cite{Jozsa1994}.


In order to understand the entanglement with the tardigrade tun, we first expand the dressed states $\ket{g_{\mathrm{BT}}}$ and $\ket{e_{\mathrm{BT}}}$ back into the larger qubit-tardigrade subspace using Eq.~(\ref{EQ_E}).
Since the states of the oscillators involved in Eq.~(\ref{EQ_E}) are orthogonal, the tardigrade is effectively modeled by a qubit. The density matrix of this three-qubit system (qubit A - qubit B - tardigrade) can then be reconstructed from the tomographic data (see SI). Using tangles based on negativity as entanglement quantifiers~\cite{CKW,Zhang2017,Ou2007}, we observe that entanglement in various bipartitions of the tripartite system as well as genuine tripartite entanglement are monotonically increasing function of $\theta$, the coupling strength between qubit B and tardigrade tun, see Fig.~\ref{fig:tomo_entanglement}. In particular, for any finite interaction strength the tardigrade is entangled with both qubits. 

The results reported here have been measured in a single experimental sequence, i.e. using one tardigrade in its tun state. Two more experiments have been conducted and we wish to point out that it is very important for the revival of the animal to change the external temperature and pressure gently. See SI for their profiles that led to the successful revival.

We conclude by revisiting Bohr's assertion on the impossibility of conducting quantum experiments with living organisms.
Our present investigation is perhaps the closest realisation combining biological matter and quantum matter available with present-day technology. While one might expect similar physical results from inanimate object with similar composition to the tardigrade, we emphasise that entanglement is observed with entire organism that retains its biological functionality post experiment. At the same time, the tardigrade survived the most extreme and prolonged conditions it has ever been exposed to, demonstrating that cryptobiosis (latent life) is truly ametabolic.
We hope this will stimulate further experiments with the states of the animal being more and more macroscopically distinguishable. Our work provides a first step in the exciting direction of creating hybrid systems consisting of living matter and quantum bits.

\bibliography{tardigrades}



\section*{Additional Information}
 This work was supported by the National Research Foundation and the Ministry of Education of Singapore, and the Polish National Agency for Academic Exchange NAWA Project No. PPN/PPO/2018/1/00007/U/00001.
The authors thank Dr. Farshad Foroughi (CNRS Grenoble) for the assistance in the fabrication of the transmon qubits used in the experiment.

{\bf Author Contrubutions} All authors researched and wrote this paper.

{\bf Supplementary Information} Supplementary information is available for this paper.

{\bf Competing Interests} The authors declare that they have no competing financial interests.

{\bf Correspondence} Correspondence and requests for materials should be addressed to rdumke@ntu.edu.sg.


\end{document}


\title{Entanglement between superconducting qubits and a tardigrade \\ Supplementary Information}

\author{K. S. Lee}
\affiliation{School of Physical and Mathematical Sciences, Nanyang Technological University, Singapore}

\author{Y. P. Tan}
\affiliation{School of Physical and Mathematical Sciences, Nanyang Technological University, Singapore}

\author{L. H. Nguyen}
\affiliation{School of Physical and Mathematical Sciences, Nanyang Technological University, Singapore}

\author{R. P. Budoyo}
\affiliation{Centre for Quantum Technologies, National University of Singapore, Singapore}

\author{K. H. Park}
\affiliation{Centre for Quantum Technologies, National University of Singapore, Singapore}

\author{C. Hufnagel}
\affiliation{Centre for Quantum Technologies, National University of Singapore, Singapore}

\author{Y.~S.~Yap}
\affiliation{Faculty of Science and Centre for Sustainable Nanomaterials (CSNano), Universiti Teknologi Malaysia, 81310 UTM Johor Bahru, Johor, Malaysia}
\affiliation{Centre for Quantum Technologies, National University of Singapore, Singapore}

\author{N. M{\o}bjerg}
\affiliation{Department of Biology, University of Copenhagen, Denmark}

\author{V. Vedral}
\affiliation{Department of Physics, University of Oxford, United Kingdom}
\affiliation{Centre for Quantum Technologies, National University of Singapore, Singapore}
\affiliation{Department of Physics, National University of Singapore, Singapore}

\author{T. Paterek}
\affiliation{Institute of Theoretical Physics and Astrophysics, University of Gda{\'n}sk, Poland}

\author{R. Dumke}
\affiliation{School of Physical and Mathematical Sciences, Nanyang Technological University, Singapore}
\affiliation{Centre for Quantum Technologies, National University of Singapore, Singapore}

\maketitle

\section{Tardigrades in the tun state}
In the tun state, the tardigrade legs are withdrawn and the body longitudinally contracted to appear more cuboid in structure, see Fig. \ref{fig:tun} below.
In our experiments, adult specimens of $\textit{Ramazzottius varieornatus}$ Bertolani and Kinchin, 1993 (Eutardigrada, Ramazzottiidae) were collected in February 2018 from a roof gutter in Niv\r{a}, Denmark (N 55$^\circ$56.685', E 12$^\circ$29.775'). The roof gutter sample, containing the tardigrades, was frozen under wet conditions and stored at $-20^\circ$C until October 2020, when the sample was thawed, diluted in ultrapure water (Millipore Milli-Q$\textsuperscript{\textregistered}$ Reference, Merck, Darmstadt, Germany) and examined for active adult tardigrades with the help of a stereomicroscope. Single adult tardigrades were removed from the sample with the aid of hand pulled glass pasteur pipettes and desiccated on filter paper substrates for use in the experiments. Subsequent attempts to mechanically remove the tardigrades from the filter paper irreversibly damage their cuticle, and hamper their ability to revive in water.

\begin{figure}[h!]
    \centering
    \includegraphics[width=.85\linewidth]{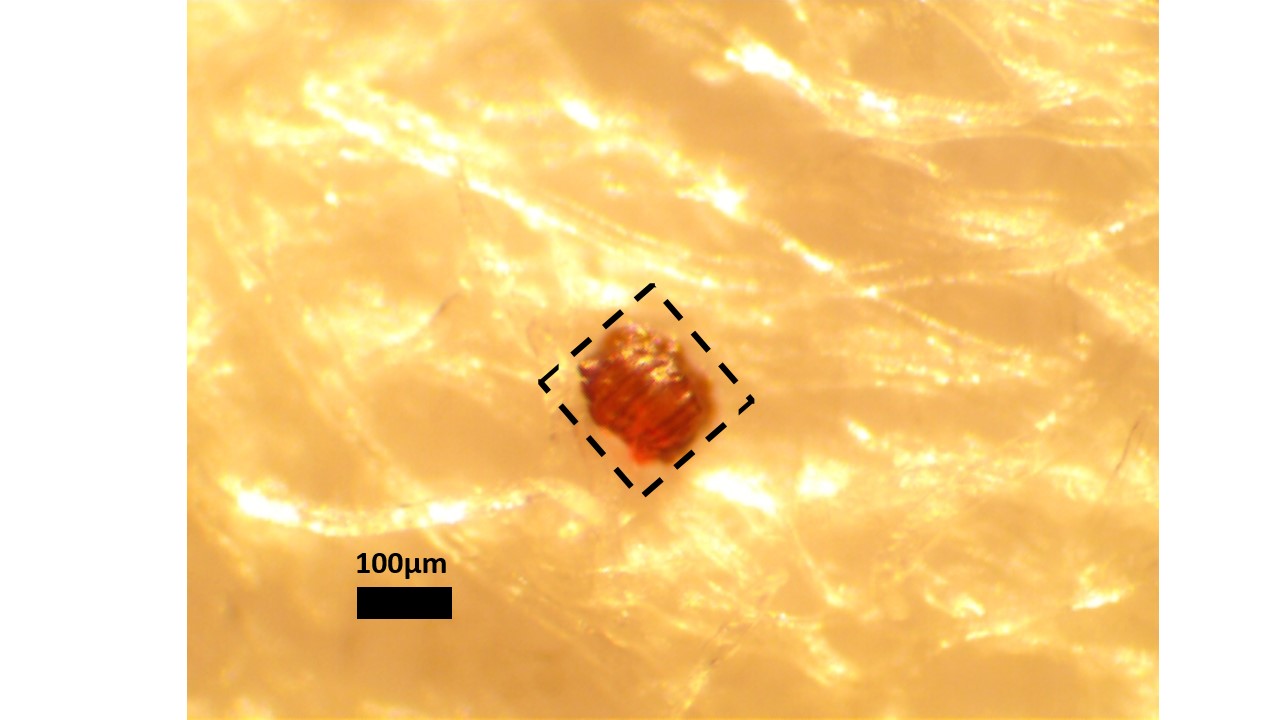}
    \caption{Tardigrade in the tun state, attached to filter paper. In the experiment, the filter paper is cut to a square shape of about 150 $\mu$m length, approximately the size of the dashed square.}
    \label{fig:tun}
\end{figure}

\section{Temperature and pressure profiles}

We have attempted three experimental sequences, each one with a different tardigrade.
A successful revival was only achieved in the third experimental trial. In this trial a slower venting process was undertaken to implement gentler pressure gradients on the warmup (room pressure was reintroduced in about 1 hour, while typically it takes only about 15 minutes). There are no pressure sensors inside the mixing chamber where the qubit and tardigrade sits. However, the sensors in the outer chambers set an upper limit on the pressures of the mixing chamber. The pressure and temperature profiles in the refrigerator for the experimental run that yielded a successful revival, from 22 Jan 2021 to 25 Jan 2021 (cooldown) and 11 Feb 2021 to 14 Feb 2021 (warmup), are shown in Fig. \ref{fig:cooldown} below. The cooldown took about 2 days to reach sub 10 mK temperatures while the pressure drop to around $6\times 10^{-3}$ mbar took about half an hour. The final pressure is likely to be much lower than $10^{-6}$ mbar \cite{Landra2019}. The warmup took about 3 days for temperature and an hour for pressure to return to room values.
\begin{figure}[h!]
    \centering
    \includegraphics[width=\linewidth]{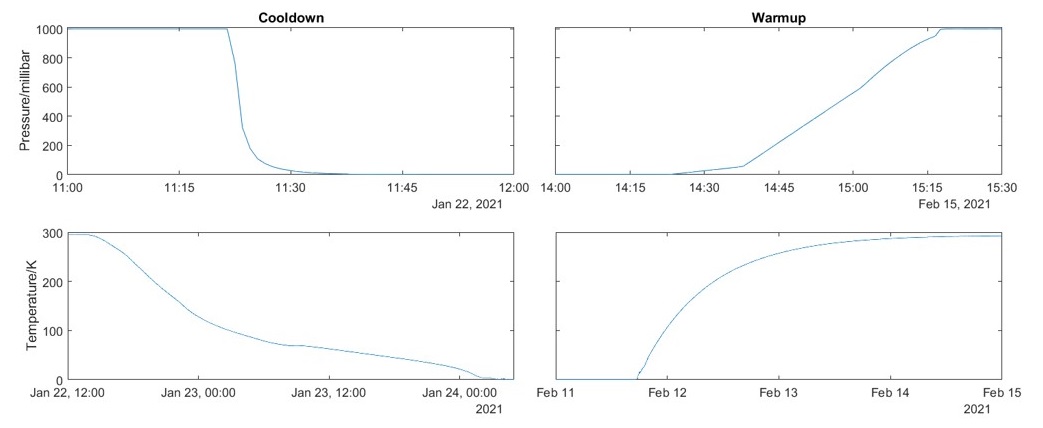}
    \caption{Left: Pressure and temperature profiles during the cooldown. Right: Pressure and temperature profiles during warmup. The timezone on the horizontal axis is Singapore Standard Time (GMT +8).}
    \label{fig:cooldown}
\end{figure}

\begin{figure}[h!]
    \centering
    \includegraphics[width=\linewidth]{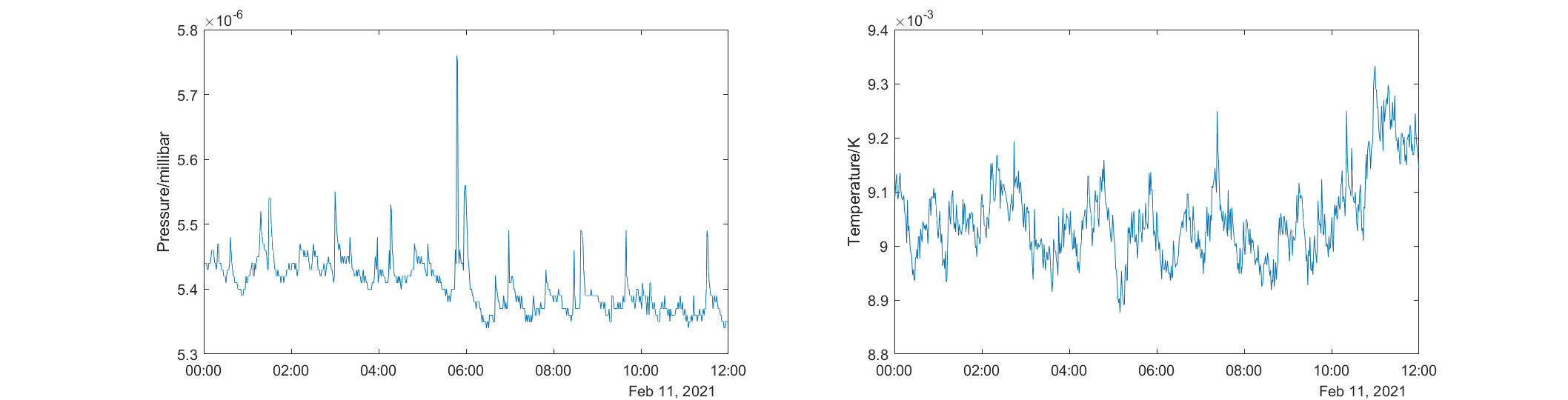}
    \caption{Fluctuations in the pressure and temperature of the refridgerator over a 12 hour period. Temperatures remain sub 10 mK, with pressures staying around $6\times10^{-6}$ mbar. The timezone on the horizontal axis is Singapore Standard Time (GMT +8).}
    \label{fig:fluctuations}
\end{figure}

After the cooldown, the pressure and temperature stabilises to around $6\times 10^{-6}$ mbar and 10 mK, with some fluctuations. A typical 12 hour period is shown in Fig. \ref{fig:fluctuations}.

\section{Reconstruction of the tripartite density matrix}

The experimentally determined $4\times4$ matrix of the tripartite system composed of qubit A, qubit B and tardigrade tun is not necessarily physical, i.e. positive semi-definite with unit trace, due to experimental errors and noise. Maximum likelihood method was used to reconstruct a physical density matrix $\rho$ closest to the experimental data, by following the procedure described in Ref.~\cite{James2001}. This density matrix is presented in the main text.
Note that it is written in the basis $\{ \ket{0g},\ket{0e}, \ket{1g}, \ket{1e} \}$,
where the combined states $\ket{g}$ and $\ket{e}$ of the qubit B - tardigrade system are used.

In order to isolate the degrees of freedom related to the tardigrade tun we rewrite the density matrix $\rho$ by expanding states $\ket{g}$ and $\ket{e}$. To this end we use Eq. (4) of the main text, which we here repeat for convenience:
\begin{eqnarray}
\ket{g} & = & \ket{0} \ket{0 \dots 0} \equiv \ket{0} \ket{\underline{0}}, \\
\ket{e} & = & \cos \frac{\theta}{2}\ket{1}\ket{0 \dots 0}+\sin \frac{\theta}{2}\ket{0}\ket{\psi_1} \equiv \cos \frac{\theta}{2}\ket{1}\ket{\underline{0}}+\sin \frac{\theta}{2}\ket{0}\ket{\underline{1}},
\end{eqnarray}
where we introduced qubit notation for the tardigrade degrees of freedom using the fact that states $\ket{0 \dots 0}$ and $\ket{\psi_1}$ are orthogonal.
For example, the operator $\ket{0g}\bra{0e}$ corresponding to the matrix element of $\rho$ is mapped to the operator in tensor product of three Hilbert spaces:
\begin{align}
    \ket{0g}\bra{0e}
    &=\cos \frac{\theta}{2}\ket{00\underline{0}}\bra{01\underline{0}}+\sin \frac{\theta}{2} \ket{00\underline{0}}\bra{00\underline{1}},
\end{align}
and so on. The expression on the right-hand side is written in the following order $\mathcal{H}_A \otimes \mathcal{H}_B \otimes \mathcal{H}_T$, where the Hilbert spaces describe qubit A, qubit B and effective qubit for the tardigrade tun, respectively. The reconstructed density matrix written in this space explicitly depends on the coupling strength $\theta$ are reads:
\begin{equation}
    \rho_{ABT} = \bordermatrix{ &\bra{000} & \bra{010} & \bra{001} & \bra{011} & \bra{100} & \bra{110} & \bra{101} & \bra{111}\cr
    &  \rho_{11} & \rho_{12} \cos \frac{\theta}{2} &  \rho_{12} \sin \frac{\theta}{2} & 0 & \rho_{13} & \rho_{14} \cos \frac{\theta}{2} &  \rho_{14} \sin \frac{\theta}{2} & 0 \cr &\rho_{21} \cos \frac{\theta}{2} & \rho_{22} \cos^2 \frac{\theta}{2} &  \rho_{22} \cos \frac{\theta}{2} \sin \frac{\theta}{2} & 0 & \rho_{23} \cos \frac{\theta}{2} & \rho_{24} \cos^2 \frac{\theta}{2} &   \rho_{24} \cos \frac{\theta}{2} \sin \frac{\theta}{2} & 0 \cr&
     \rho_{21} \sin \frac{\theta}{2} &  \rho_{22} \cos \frac{\theta}{2} \sin \frac{\theta}{2} & \rho_{22} \sin^2 \frac{\theta}{2} & 0 &  \rho_{23} \sin \frac{\theta}{2} &  \rho_{24} \cos \frac{\theta}{2} \sin \frac{\theta}{2} & \rho_{24} \sin^2 \frac{\theta}{2} & 0 \cr&
    0 & 0 & 0 & 0 & 0 & 0 & 0 & 0 \cr&
    \rho_{31} & \rho_{32} \cos \frac{\theta}{2} &  \rho_{32} \sin \frac{\theta}{2} & 0 & \rho_{33} & \rho_{34} \cos \frac{\theta}{2} &  \rho_{34} \sin \frac{\theta}{2} & 0 \cr&
    \rho_{41} \cos \frac{\theta}{2} & \rho_{42} \cos^2 \frac{\theta}{2} &  \rho_{42} \cos \frac{\theta}{2} \sin \frac{\theta}{2} & 0 & \rho_{43} \cos \frac{\theta}{2} & \rho_{44} \cos^2 \frac{\theta}{2} &  \rho_{44} \cos \frac{\theta}{2} \sin \frac{\theta}{2} & 0 \cr&
    \rho_{41} \sin \frac{\theta}{2} &  \rho_{42} \cos \frac{\theta}{2} \sin \frac{\theta}{2} & \rho_{42} \sin^2 \frac{\theta}{2} & 0 &  \rho_{43} \sin \frac{\theta}{2} & \rho_{44} \cos \frac{\theta}{2} \sin \frac{\theta}{2} & \rho_{44} \sin^2 \frac{\theta}{2} & 0 \cr&
    0 & 0 & 0 & 0 & 0 & 0 & 0 & 0},
\end{equation}

where $\rho_{ij}$ are the matrix elements of $\rho$. Note that $\rho_{ABT}$ admits 28 zero values reflecting the assumption that the system is not energetic enough for simultaneous excitation of qubit B and the tardigrade, i.e. terms with $\ket{01\underline{1}},\ket{11\underline{1}},\bra{01\underline{1}},\bra{11\underline{1}}$ have vanishing corresponding matrix elements.
The density matrix $\rho_{ABT}$ was used to compute entanglement presented in the main text.

\section{Entanglement quantifiers}

To quantify various types of entanglement in the tripartite system, we utilise so-called tangles previously defined in \cite{Zhang2017}. 
Let us begin with (doubled) negativity~\cite{vidal2002} being well-known quantifier of bipartite entanglement:
\begin{equation}
    N_{X:Y}=|| \rho_{XY}^{T_X}||-1,
\end{equation}
where $||M|| = \Tr \sqrt{M^\dagger M}$ is the trace norm of matrix $M$ and $T_X$ stands for the partial transpose over subsystem X. 
These quantities for various bipartitions involving the tardigrade are plotted in Fig. 3c of the main text.

In order to quantify genuine tripartite entanglement, i.e. entanglement irreducible to pairwise entanglements, we compute so-called $\pi$-tangle as follows:
\begin{equation}
    \pi = \frac{\pi_A+\pi_B+\pi_C}{3},
\end{equation}
where
\begin{align}
    \pi_A&=N_{A(BC)}^2 -N_{AB}^2-N_{AC}^2,\\
    \pi_B&=N_{B(AC)}^2 -N_{BA}^2-N_{BC}^2,\\
    \pi_C&=N_{C(AB)}^2 -N_{CA}^2-N_{CB}^2.
\end{align}
Fig. 3c of the main text shows that this form of entanglement is present for any non-zero coupling strength $\theta$.

\section{Dielectric model}
The transmon is an anharmonic quantum oscillator with the Hamiltonian,
\begin{equation}
    H_{\mathrm{tr}} = \hbar \sum_j \omega_j \ket{j}\bra{j},
\end{equation}
where
\begin{equation}
    \omega_j=(\omega -\frac{\delta}{2})j+\frac{\delta}{2}j^2,
\end{equation}
where $\delta$ is the anharmonicity and $\hbar \omega=\sqrt{E_cE_j}-E_c$, with $E_c=\frac{e^2}{2C}$ being the energy of the shunt capacitor with capacitance $C$, and $E_j$ is the Josephson energy associated with the Josephson junction.

By using the transmon as a qubit, we restrict ourselves only to the ground and first excited state, allowing us to neglect the anharmonicity and rewrite the Hamiltonian reduced to this subspace as~\cite{Krantz2019}:
\begin{equation}
    H_q = \frac{\omega}{2}\sigma_z.
\end{equation}

Due to the geometry of the transmon qubit, the presence of the tardigrade tun non-trivially increases the relative permittivity across the shunt capacitor, $\epsilon_r$. The relative permittivity was modeled using finite element method on ANSYS Maxwell. The altered capacitance, $C'=\epsilon_r C$, then lowers the charging energy and consequently the qubit frequency.

\bibliography{tardigrades}